# ShIFT: A Semi-haptic Interface for Flute Tutoring

Gus G. Xia
New York University, Shanghai
gxia@nyu.edu

Carter O. Jacobsen
Qiawen Chen
Xing-Dong Yang
Dartmouth College
{carter.o.Jacobsen.19, qianwen.chen.gr,
xing-dong.yang}@dartmouth.edu

Roger B. Dannenberg
Carnegie Mellon University
rbd@cs.cmu.edu

## ABSTRACT
Traditional instrument learning is time-consuming; it begins with learning music notation and necessitates layers of sophistication and abstraction. *Haptic interfaces* open another door to the music world for the vast majority of beginners when traditional training methods are not effective. However, existing haptic interfaces can only deal with specially designed pieces with great restrictions on performance duration and pitch range due to the fact that not all performance motions could be guided haptically for most instruments. Our *ShIFT* system breaks such restrictions using a *semi-haptic* interface. For the first time, the pitch range of the haptically learned pieces goes beyond an octave (with the fingering motion covers most of the possible choices) and the duration of learned pieces cover a whole phrase. This significant change leads to a more realistic instrument learning process. Experiments show that our semi-haptic interface is effective as long as learners are not "tone deaf." Using our prototype device, the learning rate is about 30% faster compared to learning from videos.

## 1. INTRODUCTION
People wish to play an instrument, even if only simple pop song for self-entertainment or *Happy Birthday* for their childrens' birthday parties. However, learning an instrument is a time-consuming process, which usually necessitates layers of sophistication and abstraction (as shown in **Figure 1**). This study aims to reduce the layers of such a complex learning procedure using a novel semi-haptic interface, leading to a more effective instrument learning method when traditional training procedures are unsuccessful.

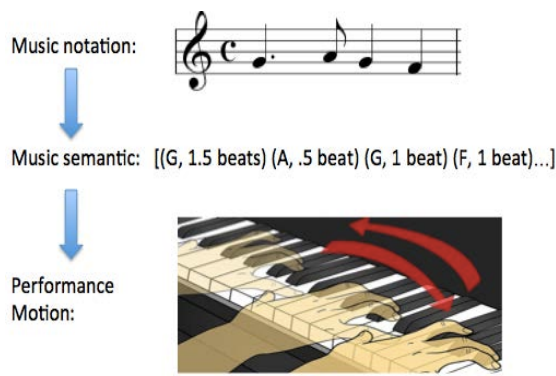

**Figure 1. A three-layers abstraction of the traditional instrument learning process.**

### 1.1 From Traditional to Haptic Learning
To learn to play a song, a layman first needs to learn *music notation*, i.e., to decode the *music semantics* (e.g., pitch and rhythm) from the visual representation of a piece of music. This process is often coupled with imagining or "singing" the tone is one's mind. The second step is to master the mapping from the music notation to *performance motion* (e.g., a common piano performance motion is to push down a key while holding that key for a certain duration). Finally, they must memorize the whole piece of music by practicing it repeatedly. All of these steps require numerous hours of practice. Though we do see exceptions of truly gifted people who can "decode" an instrument in minutes by their sharp ears and great motor sense, the vast majority of beginners suffer to learn even the basics and many give up in the process [13].

As we can see from Figure 1, though the memorization process usually involves all three layers of abstraction of a piece of music (i.e., to learn music notation, music semantics, and performance motion sequences altogether), it is *sufficient* to reproduce a piece using only the correct performance motion sequence. Also, compared to music notation and semantics, which are "abstract representations," performance motion is more of a "concrete behavior" that can be learned haptically and reproduced by muscle memory. This insight motivates us to teach instruments with a haptic interface, skipping other layers of abstraction in order to achieve a much faster and less painful music learning process.

### 1.2 The Choice of a *Semi*-haptic Interface
We choose to use a *tin whistle* (as shown in Figure 2), a type of vertical flute, as the learning instrument. Like other types of wind instruments, its performance motion consists of two inter-related components: *fingering* and *breathing*. The former refers to covering the correct holes of a flute, while the latter means to blow in the flute at a proper strength. Our semi-haptic interface *only* guides the fingering haptically, assuming that the learners can figure out the proper breathing by exploring the intrinsic finger-breath relationship.

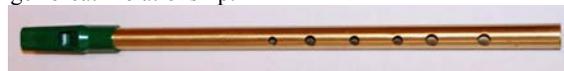

**Figure 2. An illustration of a tin whistle.**

A tricky while common feature of wind instruments is "one-to-many mapping" between finger positions and pitches. For example, for a C4 flute with all the six holes covered, a relatively *soft* breath will produce C4, while a relatively *hard* breath will produce C5. If the breath is too soft or too hard, the flute will not make any proper sound. In other words, to perform each note (pitch in the right octave) using the flute requires a particular combination of fingering and breathing.

We see that fingering and breathing control the flute in very different ways. While it is straightforward to define a correct finger position, to define "proper" breathing is much more subjective because one user's soft breath could be another's strong breath. Also, compared with different fingering positions, the number of output pitches influenced by breathing is limited, so learners are more likely to figure out the correct breathing as long as they can tell the difference between octaves. Therefore, we chose to guide only the complex fingering, i.e., to guide finger motions with a computer-controlled interface, and let the learners to explore their own proper breathing.

### 1.3 An Overview of Experimental Design

We evaluated the effectiveness of the proposed haptic semi-guided learning method step-by-step in a three-phase experiment. In the first phase, the task is to reproduce *only* the correct fingering sequence after learning a piece of music haptically with synchronized audio playback. We observed that for an 8-measure piece that contains a complete phrase, most subjects achieved the task within 30 minutes. This result indicates that a haptic interface is effective for learning flute fingering, i.e., learners are able to skip the abstract music representations (of music notation and semantics) and directly learn the fingering motions haptically using our prototype device.

In the second phase, we tasked subjects with performing a correct pitch sequence. This means not only to reproduce the correct fingering sequence but also to figure out the proper breathing for each note. We observed that for an 8-measure piece with similar difficulty to the piece used in the first phase, most subjects achieved the task within 30 minutes using our semi-haptic interface with audio playback, unless they claimed themselves as "tone deaf" and could not distinguish pitch differences in octaves. This result indicates that *semi-haptic* guidance is a valid method for most learners, i.e., learners are able to reproduce the full performance motion under semi-haptic guidance by exploring the intrinsic finger-breath relationship. Another observation is that most learners, by the time they learned to reproduce the correct pitch sequence, were *not* able to hum the melody to the same extent. This result indicates that memorizing performance motions by haptic guidance is generally faster compared to memorizing music semantics by audio playback.

In the third phase we tasked participants with learning to play two songs of similar difficulty in two ways: one learned from a video (with audio playback) showing detailed fingering using an unmodified flute, and another learned haptically using our prototype as in the second phase study. We decided to keep the length of the learned pieces as we observed that subjects started to lose their patience after 30 minutes of Lab study. The experiment shows that the learning rate associated with semi-haptic guidance is about 30% faster compared with learning from video. This result indicates that the proposed semi-haptic interface, besides saving time from memorizing music notation and semantics, is significantly more effective in memorizing performance sequences compared with learning visually from videos.

### 2. RELATED WORK

Many studies have explored the effects of haptic guidance for motor skill learning, including [1]-[4], [11], [12], [14]. More recently, we saw haptic interfaces being applied to learn music instruments. Grindlay [7] applied haptic guidance to learn drum kick sequences, in which the subjects learned the sequences under three conditions: audio (only), haptic (only), and audio-haptic. Experiments showed that audio-haptic guidance is the best. This guidance, on average, shrank the error by 18% compared with just learning from audio. Huang et al. [8] developed a wearable tactile device that looks like a glove to learn short piano segments passively, in which the subjects learned the sequence actively just once and then reinforced the learned piece passively (while actively doing some reading tasks) under two conditions: audio (only) and audio-haptic. Experiments showed that audio-haptic guidance is a better passive learning strategy; after 30 minutes of passive learning, the mean improvement of the haptic group was 3.44 notes while the improvement of the audio group was negative. Fujii et al. [5] developed a haptic device to learn the Theremin and compared the learning process under three conditions: visual, haptic, and visual-haptic. Though quantitative results were not reported, we see that with haptic guidance the learned motion traces of beginners match better to the traces of expert performers compared to the baseline.

Despite these efforts of applying haptic interfaces to instrument learning, we see a fundamental limitation: only a small portion of performance motion is learned. As a consequence, learners can only learn specially designed pieces with great restriction on pitch ranges and piece duration. (Only 5 pitches were learned in [8] and [5], while only the kick motion of a single drum, rather than the performance motion of an actual drum set, was learned in [7].) The reasons for such restrictions are twofold. First, to learn only a limited portion of performance motion leads to a faster learning process, which is a good choice for pilot studies. Second, and more importantly, it is only feasible to guide a *part* of the performance motion haptically for most instruments. We have already discussed in the introduction that it is not feasible to guide breathing haptically because different people have very different breathing strengths. Similarly, different learners have different arm lengths; therefore, it is not a good choice to haptically guide arm motion (which is an essential part of performance) across a keyboard or a drum set.

Compared with previous studies, this study makes a significant step towards a working system for instrument learning in the real world. Thanks to the carefully designed *semi-haptic* interface, the durations of the learned songs become significantly longer and the pitch ranges become much wider compared to the pieces created in previous studies. For the first time, the learned pitch range goes beyond an octave and the learned fingering motion covers most of the possible choices (leaving out only the half holes and ultra-high notes which require advanced techniques.)

### 3. SYSTEM COMPONENTS

To achieve the guidance, we use servos to push or pull fingers into positioni. We attached 6 servomotors to the underside of the whistle (as shown in Figure 3). The cylindrical shape of the flute allowed for two 3D printed sleeves to slide onto the flute and then be tightened down once positioned correctly. Two sleeves were used to accommodate the user's hand position by placing the securing portion of each sleeve on the side of the flute opposite the hand (left hand on the top three holes and and right hand on the bottom three). In the current prototype the sleeves were secured to the flute by superglue, but it is feasible that the fastening could be clamped or pinned such that the entire device could be removed from the flute and reattached. Each sleeve has mounting slots for a motor. The upper sleeve holds its motors vertically, with the intent that users will use their left thumbs above the motors to support the flute. The bottom sleeve positions the 4[th] and 6[th] motors vertically while the 5[th] is horizontal to allow users to rest their right thumbs on this motor casing.

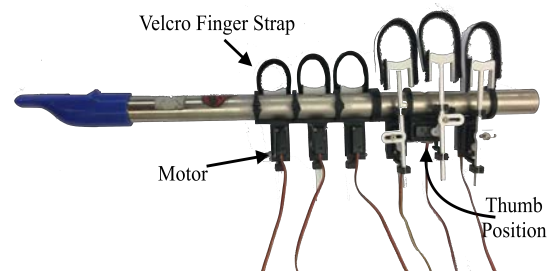

**Figure 3. An illustration of the haptic guidance hardware.**

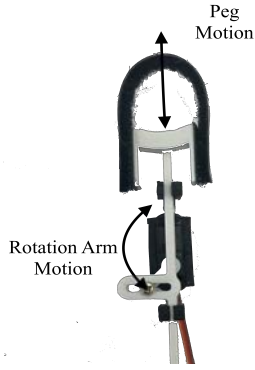

Figure 4. An illustration of a single motor and peg.

Attached to the motors are guide pieces for the finger pegs. These guide fixtures have two arms each where the pegs slot in. Each peg, when not attached to the motor, is able to move freely up and down alongside the device. The peg itself contains a cradle designed to sit just past the user's first knuckle. Having the cradle point in this position allows the user to adjust their finger pad to cover the hole more fully. Attached to the sides of the peg is a Velcro strap, which can be adjusted to the necessary length to accommodate all finger sizes. Near the middle of the peg, sitting between the guide posts, is a slot protruding from the side of the peg.

Through this slot in the peg, a screw is fed through that can move freely by itself. This screw is attached to an arm of the servo motor. We use $135°$ of the servo motor rotation in a scotch yoke mechanism shown in Figure 4. This system worked well in practice, however due to many connections, indirect force, and very small motors, it is imperative that users relax their fingers almost completely while using the device.

The motors are connected to an Arduino Due for signals and the entire set of motors uses an independent power supply rather than being powered from the Arduino. The signals are sent through any computer connected to the Arduino by way of a simple UI where the user can trigger each motor to move its peg to high or low individually or select pre-transcribed songs to play. When songs are selected, the motors will move synchronously with the song being played through the computer's speakers.

## 4. EXPERIMENTS
## 4.1 Study 1: Evaluation of Haptic Interface

The first experiment is to test whether people can learn finger motions haptically after a learning period with repeated haptic guidance. Both the learning and testing phases involve synthesized audio playback synchronized with real-time finger motions so that the learners do not have to blow the flute to make any sound.

As in previous studies [5][7][8], we chose the criterion to be the correct note sequence (as in Study2&3). The reasons are twofold. First, a correct note sequence serves as a first order approximation of a correct performance. Second, note sequence is a more quantitative measurement compared to note durations and dynamics, both of which could vary a lot between different interpretations [15] and hence are much more difficult to evaluate.

### 4.1.1 The Music to Learn
We composed an 8-bar piece based on the first two phrases of a famous Irish folk song named *Sally Garden* [6]. By modifying existing songs, we keep the learning materials more realistic while avoiding the familiarity of the song by any subjects before the experiment. Figure 5 shows the score. Only for the first experiment, we intentionally constrain the pitch range within an octave, in which case the finger motions and pitches have a nice one-to-one correspondence so that the synthesizer can decide the pitch to be played purely according to the fingering.

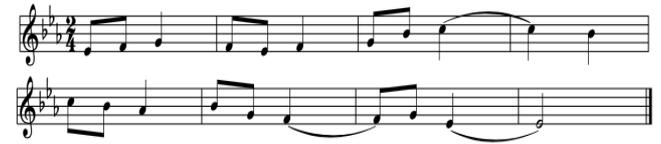

Figure 5. The score for the first experiment.

Table 1 shows some basic statistics of the finger motion of adjacent note pairs, which approximately reveals the learning difficulty of the piece. Here, the first column represents the number of moving fingers and the second column shows the count of these instances. For example, the $3^{rd}$ row that starts with 2 means that there are 3 times where users should move 2 fingers at the same time. Larger intervals and motions that involve more fingers usually lead to a more difficult piece of music [9]. Therefore, this piece is relatively easy since we see smaller numbers associated with larger finger movements.

Table 1. Basic statistics of the adjacent note pairs, which approximately reveal the learning difficulty of the $1^{st}$ piece.

| No. of moving fingers | Count |
|---|---|
| 1 | 12 |
| 2 | 3 |
| 5 | 2 |

### 4.1.2 Participants
Sixteen paid participants (7 males and 9 females) between the age of 21 and 35 participated in the study. All participants had no experience playing flute and reported no familiarity with the composed piece.

### 4.1.3 Task and Procedure
The task consists of two parts: *learning* and *testing*. The learning part required the participants to wear the device and feel the guided finger motions using the haptic interface while listening to synchronized audio playback. Participants try to memorize the motion sequence during repeated guidance. The testing part required the participants to reproduce the learned finger motion sequence on another flute (without the haptic component). The participants started with the learning mode and could switch to the testing mode if they were confident enough or simply wanted to have a try. If participants failed in the testing, they were free to switch back to the learning mode unless they decided to give up the task. Participants were asked to finish the task as fast and as accurately as possible, and the task is marked *complete* upon the first successful reproduction of the learned sequence.

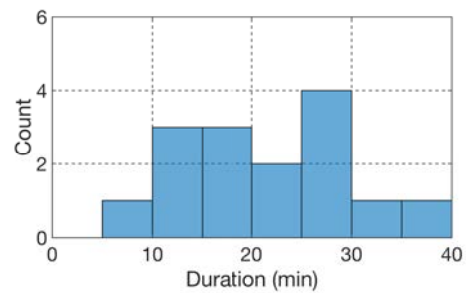

Figure 6. The result of the $1^{st}$ experiment.

### 4.1.4 Results and Discussion

We recorded the total duration each participant spent to complete the task as the main indicator of learning rate. 15 out of the 16 participants completed the task. (One participant gave up because her hands were sweating so much, she could not even properly hold the flute). Figure 6 shows a histogram of the learning durations, where we see that most participants finished the task within 30 minutes. This result indicates that haptic guidance is a valid method to learn flute finger motions using the designed haptic interface.

## 4.2 Study 2: Evaluation of Semi-haptic

The second experiment is to evaluate the semi-haptic idea, i.e., to test whether subjects can learn to perform a piece with only the finger motions guided haptically. Therefore, the learning task is to not only to memorize the fingering but also to figure out the proper breathing from mistakes and the intrinsic finger-breath relationship. Unlike the first experiment, only the learning phase involved synthesized audio playback; for the testing phase, sound was made by participants' own performance.

### 4.2.1 The Music to Learn

We composed a new piece by modifying many intervals of the first piece while keeping its main pitch contours. Figure 7 shows the score. Compared to the first piece (shown in Figure 5), the fundamental difference is that its pitch range goes beyond an octave. For example, in the second system, the interval between the first note (F5) and the 7$^{th}$ note (F4) is an octave. They share the same finger position and can only be performed correctly with proper breath control.

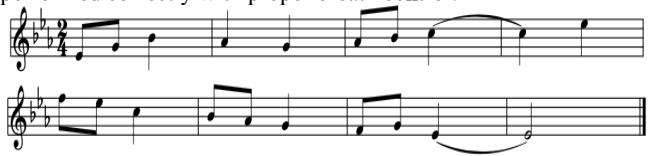

**Figure 7. The score for the second experiment.**

Table 2 reveals the difficulty of the second piece following the same format of Table 1. We can see they are exactly the same, which means these two pieces are of similar difficulty.

**Table 2. Basic statistics of the adjacent note pairs, which approximately reveal the learning difficulty of the 2$^{nd}$ piece.**

| No. of moving fingers | Count |
|---|---|
| 1 | 12 |
| 2 | 3 |
| 5 | 2 |

### 4.2.2 Participants

Sixteen paid participants (6 males and 10 females) between the age of 21 and 35 participated in the study. Two of participants overlapped with the first experiment and all other participants had no experience playing the flute. They all reported no familiarity with the composed piece.

### 4.2.3 Task and Procedure

The task consisted of three parts: pre-training, learning, and testing. In the pre-training part, we taught participants how to play a basic scale on the flute through the range of the song they would be playing, to ensure they had the baseline ability to play. After they successfully played the basic scale, we continued the experiment with the learning and testing parts as we did in the first experiment. Again, participants were asked to do their best in terms of learning rate and accuracy, and the task is completed upon the first accurate reproduction of the piece.

### 4.2.4 Results and Discussion

15 out of the 16 participants completed the task. One participant gave up because he could not distinguish the difference between Eb4 and Eb5; he claimed himself to be "tone deaf." Figure 8 shows the time taken for the participants to learn the song and play it from memory. (The time taken in the pre-training step is not included.) We see that most participants finished the task within 30 minutes as in the first experiment. This result indicates that semi-haptic guidance is a valid method for flute tutoring.

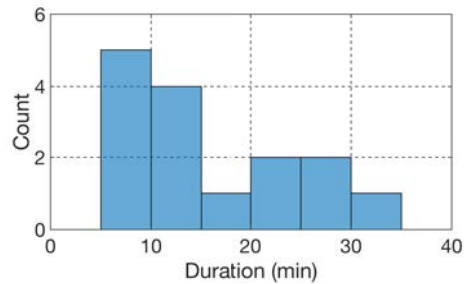

**Figure 8. The result of the 2$^{nd}$ experiment.**

Right after the participants finished the task, we asked them to hum the melody (as best they could) to see if they memorized the music itself. Interestingly, only 3 participants were able to hum the melody with a correct pitch sequence.

## 4.3 Study 3: Semi-haptic vs. Visual

The third phase study examines whether people can learn to perform a piece using the proposed semi-haptic interface or visual guidance faster. (A similar comparison between haptic and visual guidance has been conducted in [10].) We chose video guidance as the comparison because it is, so far, the best alternative to traditional music training process. Similar to haptic guidance, learning from videos also does not require much knowledge in music notation.

Half of this phase builds directly from the second phase of the experiment, as subjects must learn one of the two songs (shown below) through the same process of haptic guidance. In this phase however, we added the portion in which subjects also were asked to learn the other of the two songs through a video performance.

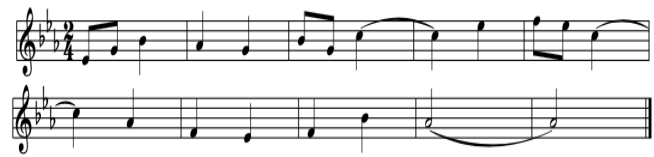

(a)    Song A.

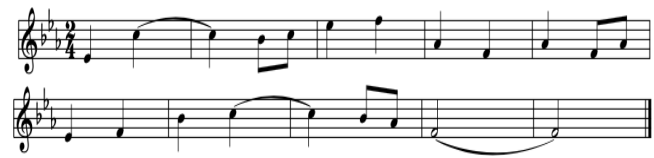

(b)    Song B.

**Figure 9. The scores for the third experiment.**

### 4.3.1 The Music to Learn

We composed two new songs of similar difficulty to each other. Figure 9 shows the scores, where the former is modified from the song used in the second experiment, and the latter is a modified Mongolian folk song, *Gada Meiren*.

To simulate a more realistic music learning experience, we made the piece to be more "jumpy" and contain larger intervals compared to the previous experiments. Table 3 show the basic statistics of finger motion for the two pieces (they two share exactly the same statistics) following the same format of Table 1 and Table 2.

**Table 3. Basic statistics of the adjacent note pairs, which approximately reveal the learning difficulty of the two pieces of 3rd experiment.**

| No. of moving fingers | Count |
|---|---|
| 1 | 7 |
| 2 | 6 |
| 3 | 2 |
| 5 | 2 |

*4.3.2 Participants*
Sixteen paid participants (11 males and 5 females between 18 and 30) took part in this phase of the study. None of them had taken part in either of the previous parts of the study and none of them had any experience on the flute. Also, they had neither heard either of the songs being used nor had the experience of learning instruments from videos.

*4.3.3 Design*
The experiment employed a 2×2 within-subject factorial design. The independent variables were *Learning method* (*semi-haptic guided, Video guided*) and *Learning piece* (*Song A, Song B*). Both were counter-balanced among participants. In other words, each participant played each of the two songs: one learned through video and the other learned through haptic guidance. We cycled through all four permutations (of song-choice and song-learning method combination) four times to produce our 16 data points. This produced 8 data points for each song and learning method combination.

*4.3.4 Task and Procedure*
Just as in the second experiment, we started by teaching subjects a scale. We then gave them the first song to learn. If the participant showed little to no progress with either of the learning methods within 30-35 minutes, we allowed them quit this portion of the study. After the first song was either marked as learned or failed, we allowed users to choose if they wanted to start the second song immediately or take a break and come back another time. If they chose to come back another time, we taught them to play a scale again before the learning phase. Both methods adopted the same learning-testing procedure as used in the first and second experiments, and the task was completed upon the first accurate reproduction of the learned pitch sequences. We also asked the participants a few concluding questions after both songs were learned.

*4.3.5 Results and Discussion*
All but three of the sixteen participants completed both songs. Of the three that showed little to no progress during one of the methods, two failed learning from video, but were successful with learning with haptic guidance, while one was successful from video and failed while learning from haptic guidance. Figure 10 shows the total durations each participant spent to complete the task using both methods, where the *x*-axis represents the time spent in haptic learning and the *y*-axis represents the time spent in video learning. We see that all but one point are above the *y*=*x* line, which shows that most participants learned the piece faster using semi-haptic guidance. Excluding the people that failed one of the methods, our method showed a statistically significant improvement (with p-value < 0.005 by pairwise t-test) and an average 30% increment in the learning rate (in terms of percentage of a piece per minute).

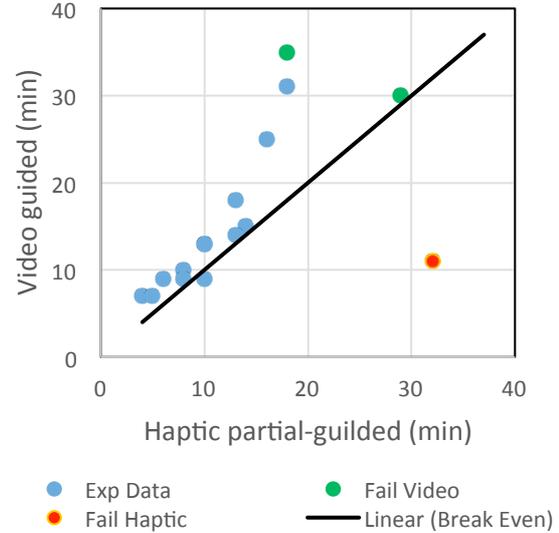

**Figure 10. The result of the 3rd experiment.**

In our post experiment interviews, the subjects that reported the most comparative assistance from the haptic guidance were those with extensive experience on similar instruments (i.e. saxophone) and those who had no musical experience at all. Those who had musical experience on non-similar instruments (i.e. piano or vocal) found similar results between the two methods. This was probably because they mainly learned through associating pitches to fingerings after hearing the song and used guessing and checking while learning rather than memorizing the fingering sequence. This result substantiates our prototype device by showing the semi-haptic approach is better than the learning from video.

## 5. CONCLUSION AND DISCUSSION

Over the past few years, we have seen promising studies that empower computer systems to better serve professional human performance. Haptic interfaces can apply artificial music intelligence in another way by letting machines take the lead in teaching humans. Following this path, we developed the *ShIFT* system, a semi-haptic interface for flute tutoring. This system breaks through the restrictions of previous haptic interfaces regarding duration, pitch, and motion ranges, achieving a more realistic instrument learning procedure. Our experiments have shown that most people are able to (at least) learn a piece as long as 8-9 bars within 30 minutes as long as they are not tone deaf. Compared to learning from videos, the learning rate is about 30% faster using our prototype device.

We see several limitations of the current device and methodology that are worth investigating further in the future. First, the device is still not strong enough and cannot be run for excessive periods of time, as the motors become overstressed, especially when users have big and strong hands. Though in the current implementation a motor can be changed on the device in under three minutes, we think it is necessary to improve the robustness of the device. Second, the current device is "position guided" and does not allow the users to violate the guided motions (e.g., if the motor spins and cause a ring to move up, there is no way for a user to push down the corresponding finger without breaking the ring). Many participants reported in the post interview that they would like to explore the motion space a little more by trial-and-error. This motivates us to build a "force guided" interface in the future, perhaps using magnetic devices. Last but not least, the focus of this paper is limited to flute beginners playing a piece of music with the correct note sequence. Learning an instrument involves much more, e.g., learning expressive dynamics and timing. It would be beneficial to see whether this interface could help professional flutists to learn complex and expressive pieces or help intermediate players to learn music notation faster.

Above all, we see this study as an important step in a long journey exploring music education using motion-guided methods.

## 6. REFERENCES

[1] T. Armstrong. 1970. *Feedback and perceptual-motor skill learning: A review of information feedback and manual guidance*. Technical Report 25, University of Michigan, Department of Psychology, Ann Arbor, MI.

[2] Y. Blandin, L. Lhuisset, and L. Proteau. 1999. *Section* A - Human Experimental Psychology. In *Cognitive processes underlying observational learning of motor skills*. Quarterly Journal of Experimental Psychology, 52(4):957–979.

[3] D. Feygin, M. Keehner, and R. Tendick. 2002. Haptic guidance: Experimental evaluation of a haptic training method for a perceptual motor skill. In *Proceedings of the 10th Symposium on Haptic Interfaces for Virtual Environment and Teleoperator Systems*, pages 40–47.

[4] P. Fitts. 1964. Perceptual-Motor Skill Learning. In *Categories of Human Learning*. New York: Academic Press.

[5] Fujii, Katsuya, et al. ACM 2015. "MoveMe: 3D haptic support for a musical instrument." In *Proceedings of the 12th International Conference on Advances in Computer Entertainment Technology*.

[6] Galway, James and P. Coulter. 1997. *Legends*. BMG Music.

[7] Grindlay, Graham. 2008. Haptic guidance benefits musical motor learning. In *HAPTICS '08: Proceedings of the 2008 Symposium on Haptic Interfaces for Virtual Environment and Teleoperator Systems, pages 397–404, Washington, DC, USA, 2008. IEEE Computer Society*.

[8] Huang, Kevin, et al. ACM, 2010. Mobile music touch: mobile tactile stimulation for passive learning. In *Proceedings of the SIGCHI conference on human factors in computing systems*.

[9] Jacobson, Jeanine Mae. 2006. *Professional piano teaching: A comprehensive piano pedagogy textbook for teaching elementary-level students*. Vol. 1. Alfred Music Publishing.

[10] M. Jones, A. Bokinsky, T. Tretter, and A. Negishi. 2005. A comparison of learning with haptic and visual modalities. In *Haptics-e*, 3(6).

[11] T. Lee, M. White, and H. Carnahan. 1990. On the role of knowledge of results in motor learning: Exporing the guidance hypothesis. *Journal of Motor Behavior*, 22(2):191–208.

[12] C. E. Lewiston. 2009. *MaGKeyS: A haptic guidance keyboard system for facilitating sensorimotor training and rehabilitation*. PhD thesis. Massachusetts Institute of Technology.

[13] Mills, Janet. 2003. Musical performance: crux or curse of music education? In *Psychology of Music* 31.3 (2003): 324-339.

[14] X.-D. Yang, W. F. Bischof, and P. Boulanger. 2008. Validating the performance of haptic motor skill training. In *HAPTICS '08: Proceedings of the 2008 Symposium on Haptic Interfaces for Virtual Environment and Teleoperator Systems*, 129–135.

[15] G. Xia. 2016 "Expressive Collaborative Music Performance via Machine Learning." CMU Ph.D. thesis..